\documentclass[twocolumn,prl,showpacs,floatfix]{revtex4}

\usepackage{subfigure}
\usepackage{graphicx}

\begin{document}

\preprint{DRAFT}
\date{January 21, 2014}

\title{Axion Dark Matter Detection using an LC Circuit}

\author{P.~Sikivie, N.~Sullivan and D.B.~Tanner}

\affiliation{Department of Physics, University of Florida, 
Gainesville, FL 32611, USA}

\begin{abstract}

It is shown that dark matter axions cause an oscillating electric 
current to flow along magnetic field lines.  The oscillating 
current induced in a strong magnetic field $\vec{B}_0$ produces 
a small magnetic field $\vec{B}_a$.  We propose to amplify 
and detect $\vec{B}_a$ using a cooled LC circuit and a very 
sensitive magnetometer.  This appears to be a suitable
approach to searching for axion dark matter in the $10^{-7}$ 
to $10^{-9}$ eV mass range.

\end{abstract}

\maketitle

Shortly after the Standard Model of elementary particles was established,
the axion was postulated \cite{axion} to explain why the strong interactions
conserve the discrete symmetries P and CP.  Further motivation for the
existence of such a particle came from the realization that cold axions
are abundantly produced during the QCD phase transition in the early
universe and that they may constitute the dark matter \cite{axdm}.
Moreover, it has been claimed recently that axions are the dark matter, 
at least in part, \cite{CABEC,case,ban} because axions form a Bose-Einstein 
condensate and this property explains the occurrence of caustic rings in 
galactic halos.  The evidence for caustic rings with the properties 
predicted by axion BEC is summarized in ref. \cite{MWhalo}.  In 
supersymmetric extensions of the Standard Model, the dark matter may be 
a mixture of axions and supersymmetric dark matter candidates \cite{Baer}.

Axion properties depend mainly on a single parameter $f_a$, called the axion
decay constant.  In particular the axion mass ($\hbar = c = 1$)
\begin{equation}
m_a \simeq 6 \cdot 10^{-6}~{\rm eV}~{10^{12}~{\rm GeV} \over f_a}
\label{mass}
\end{equation}
and its coupling to two photons
\begin{equation}
{\cal L}_{a\gamma\gamma} =
- g~a(x) \vec{E}(x)\cdot\vec{B}(x)
\label{emcoupl}
\end{equation}
with $g = g_\gamma {\alpha \over \pi f_a}$.  Here $a(x)$ is the axion field, 
$\vec{E}(x)$ and $\vec{B}(x)$ the electric and magnetic fields, $\alpha$ the 
fine structure constant, and $g_\gamma$ a model-dependent coefficient of order 
one.  $g_\gamma \simeq - 0.97$ in the KSVZ model \cite{KSVZ} whereas $g_\gamma 
\simeq 0.36$ in the DFSZ model \cite{DFSZ}.  Cold axions are produced during 
the QCD phase transition, when the axion mass turns on and the axion field 
begins to oscillate in response. The resulting axion cosmological energy 
density is proportional to $(f_a)^{7 \over 6}$ and, in the simplest case, 
reaches the critical energy density for closing the universe when $f_a$ is 
of order 10$^{12}$ GeV \cite{axdm}.  This suggests that the most promising 
mass range in axion searches is near 10$^{-5}$ eV.  This happens to be 
approximately where the cavity axion detection technique \cite{axdet} is 
most feasible and where the ADMX experiment \cite{ADMX} is searching at 
present.

However, it is desirable to search for axion dark matter over the 
widest possible mass range because the axion mass is, in reality, 
poorly constrained. In particular, it has been argued that if there is 
no inflation after the Peccei-Quinn phase transition, the contribution 
of axion strings to the axion cosmological energy density \cite{axcos} 
implies that the preferred mass for dark matter axions is in the $10^{-3}$ 
to $10^{-4}$ eV mass range \cite{EPS}. On the other hand, if there is 
inflation after the Peccei-Quinn phase transition, the axion field gets 
homogenized during inflation and the homogenized field may accidentally 
lie close to the minimum of its effective potential \cite{sypi}, in which 
case axions may be the dark matter for masses much smaller than 10$^{-5}$ eV.  
String theory favors values of $f_a$ near the Planck scale and hence very 
small axion masses \cite{strax}.  It also predicts a variety of axion-like 
particles (ALPs) in addition to the axion that solves the strong CP problem
\cite{Svrcek}. For such ALPs there is no general relationship between the 
coupling $g$ to two photons and the mass $m_a$.   ALPs produced by vacuum 
realignment are a form of cold dark matter with properties similar to axions 
\cite{Arias}.  The evidence for axion dark matter from axion Bose-Einstein
condensation and the phenomenology of caustic rings does not depend sharply 
on the axion or ALP mass and therefore does not tell us anything precise 
about this parameter.

Other methods aside from the cavity technique have been proposed 
to search for dark matter axions.  One proposed method consists of 
embedding an array of superconducting wires in a material transparent 
to microwave photons \cite{PDY}.  Dark matter axions convert to photons 
in the inhomogeneous magnetic field sourced by currents in the wires. 
This method appears best suited to searches for axions in the $10^{-4}$ 
eV mass range and above.  Recent papers \cite{NMR} propose the application 
of NMR techniques to axion detection.  A sample of spin polarized material 
acquires a small oscillating transverse polarization as result of the axion 
dark matter background.  The NMR techniques rely on the coupling of axions 
to nucleons.  They are best suited to searches for axion dark matter with 
masses of order $10^{-8}$ eV and below.  In addition to axion dark matter 
searches, there are searches for axions emitted by the Sun \cite{helio} and 
`shining light through the wall' experiments that attempt to produce and 
detect axions in the laboratory \cite{SLW}.  Stimulated by ref. \cite{NMR}, 
we propose here a new method to search for dark matter axions.  It exploits 
the coupling of the axion to two photons and appears suitable to axion dark 
matter searches in the $10^{-7}$ eV range and below.  Using a combination of 
the various approaches it may be possible to search for dark matter axions 
over a wide mass range, from approximately $10^{-9}$ to $10^{-4}$ eV.

The coupling of the axion to two photons, Eq.~(\ref{emcoupl}), 
implies that the inhomogenous Maxwell equations are modified 
\cite{axdet} as follows:
\begin{eqnarray}
\vec{\nabla}\cdot\vec{E} &=& 
g \vec{B}\cdot\vec{\nabla}a + \rho_{\rm el}\nonumber\\
\vec{\nabla}\times\vec{B} - {\partial \vec{E} \over \partial t} &=&
g(\vec{E}\times\vec{\nabla}a - \vec{B}{\partial a \over \partial t}) 
+ \vec{j}_{\rm el}
\label{modMax}
\end{eqnarray}
where $\rho_{\rm el}$ and $\vec{j}_{\rm el}$ are electric charge and 
current densities associated with ordinary matter. Eq. ~(\ref{modMax})
shows that, in the presence of an externally applied magnetic field 
$\vec{B}_0$, dark matter axions produce an electric current density 
$\vec{j}_a = - g \vec{B}_0 \dot{a}$, where 
$\dot{a} \equiv {\partial a \over \partial t}$.  Assuming the
magnetic field to be static, $\vec{j}_a$ oscillates with frequency
\begin{equation}
\omega = m_a (1 + {1 \over 2} \vec{v}\cdot\vec{v})
\label{axfreq}
\end{equation}
where $\vec{v}$ is the axion velocity.  Let us assume that the 
spatial extent of the externally applied magnetic field is much 
less than $m_a^{-1}$.  $\vec{j}_a$ produces then a magnetic field 
$\vec{B}_a$ such that $\vec{\nabla}\times\vec{B}_a = \vec{j}_a$.  
Our proposal is to amplify $\vec{B}_a$ using an LC circuit and 
detect the amplified field using a SQUID or SERF magnetometer.

\begin{figure}
\includegraphics[width=0.9\columnwidth]{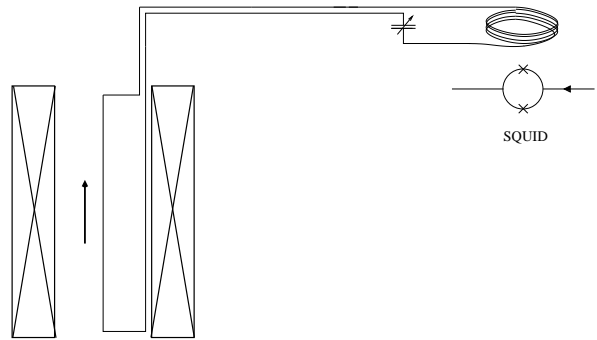}
\caption{Schematic drawing of the proposed axion dark matter
detector, in the case where the magnet is a solenoid.  The two 
crossed rectangles indicate cross-sections of the solenoid's 
windings.  The direction of the magnetic field ($\vec{B}_0$) 
produced by the solenoid is indicated by an arrow.}
\label{sch}
\end{figure}

Fig. \ref{sch} shows a schematic drawing in case the magnet producing 
$\vec{B}_0$ is a solenoid.  The field $\vec{B}_a$ has flux $\Phi_a$ through 
a LC circuit, made of superconducting wire.  Because the wire is superconducting, 
the total magnetic flux through the circuit is constant.  In the limit where the 
capacitance of the LC circuit is infinite (or the capacitor is removed), the 
current in the wire is $I = - \Phi_a/L$ where $L$ is the inductance of the 
circuit in its environment, i.e. including the effect of mutual inductances 
with neighboring circuits.  The magnetic field seen by the magnetometer is 
($\mu_0$ = 1) 
\begin{equation} 
B_d \simeq {N_d \over 2 r_d} I = - {N_d \over 2 r_d~L} \Phi_a 
\label{detf} 
\end{equation} 
where $N_d$ is the number of turns and $r_d$ the radius of the small 
coil facing the magnetometer.  Ignoring for the moment mutual inductances
with neighboring circuits, $L$ is a sum 
\begin{equation} 
L \simeq L_m + L_c + L_d 
\label{Lsum} 
\end{equation} 
of contributions $L_m$ from the large pickup loop inside the externally 
applied magnetic field, $L_d$ from the small coil facing the magnetometer, 
and $L_c$ from the co-axial cable in between. We have 
\begin{equation}
L_d = r_d N_d^2 c_d 
\label{dind} 
\end{equation} 
with
\begin{equation} 
c_d \simeq \ln\left({8 r_d \over a_d}\right) - 2 
\label{CD}
\end{equation} 
where $a_d$ is the radius of the wire in the small coil.  If mutual 
inductances are important, their effect upon $L$ must be included 
and Eq~(\ref{Lsum}) modified.  For example, if there is a single 
neighboring circuit with self-inductance $L_{22}$ and mutual inductance 
$L_{12}$ with the LC circuit, and if $\vec{B}_a$ has no flux through 
this second circuit, then
\begin{equation} 
L \simeq L_m + L_c + L_d - (L_{12})^2/L_{22}~~~\ . 
\label{comL} 
\end{equation} 
We note that the currents in the coil sourcing the $\vec{B}_0$ field are 
generally perpendicular to the currents flowing in the pickup loop, so that 
the mutual inductance between the coil and pickup loop is suppressed.  Also, 
when Eq.~(\ref{comL}) is valid, $L$ is smaller than in the $L_{12} = 0$ 
case, and hence $B_d$ is increased.  When discussing the LC circuit's
optimization and estimating the detector's sensitivity below, we will 
ignore mutual inductances.  Mutual inductances should be measured in 
any actual setup and the optimization and sensitivity estimates adjusted 
accordingly.

For finite $C$, the LC circuit resonates at frequency $\omega = 1/\sqrt{LC}$.  
When $\omega$ equals the axion rest mass, the magnitude of the current in the 
wire is multiplied by the quality factor $Q$ of the circuit and hence
\begin{equation}
B_d \simeq {Q N_d \Phi_a \over 2 L r_d} ~~~\ .
\label{fbd}
\end{equation}
We expect that a quality factor $Q$ of order $10^4$ may be achieved
by using high $T_c$ superconducting wire for the part of the LC circuit 
in the high magnetic field region \cite{HTC} and by placing superconducting 
sleeves between the LC circuit and nearby ordinary metals.  

Let us consider the case where the externally applied magnetic field is 
homogeneous, $\vec{B}_0 = B_0 \hat{z}$, as is approximately true inside 
a long solenoid.  In such a region
\begin{equation}
\vec{B}_a = - {1 \over 2} g \dot{a} B_0 \rho \hat{\phi}
\label{Ba}
\end{equation}
where ($z$, $\rho$, $\phi$) are cylindrical coordinates and $\hat{\phi}$
is the unit vector in the direction of increasing $\phi$. For the pickup 
loop depicted in Fig. 1, a rectangle whose sides $l_m$ and $r_m$ are 
approximately the length and radius of the magnet bore, the flux of 
$\vec{B}_a$ through the pickup loop is 
\begin{equation}
\Phi_a = - V_m g \dot{a} B_0
\label{genflux}
\end{equation}
with $V_m = {1 \over 4} l_m r_m^2$.  The self-inductance of the pickup 
loop is $L_m \simeq {1 \over \pi} l_m \ln\left({r_m \over a_m}\right)$ 
where $a_m$ is the radius of the wire.  We may also consider the case 
$\vec{B}_0 = B_0(\rho) \hat{\phi}$, as is approximately true in a 
toroidal magnet.  Here one introduces a circular pickup loop at 
$\rho = R_m$.  We have then Eq.~(\ref{genflux}) with
\begin{equation}
V_m B_0 = 2 \pi \int_0^{R_m} \rho~d\rho~
\int_\rho^\infty d\rho^\prime B_0(\rho^\prime)
\label{torvm}
\end{equation}
and $L_m \simeq R_m [\ln\left({8 R_m \over a_m}\right) - 2]$.

The time derivative of the axion field is related to the axion
density by $\rho_a = {1 \over 2} \dot{a}^2$.  Hence, combining 
Eqs.~(\ref{fbd}) and (\ref{genflux}), we have 
\begin{eqnarray}
B_d &\simeq& {N_d Q \over 2 r_d L} V_m g \sqrt{2 \rho_a} B_0
= 1.25 \cdot 10^{-15}~{\rm T}\cdot\nonumber\\
&\cdot&
\left({\rho_a \over {\rm GeV/cm}^3}\right)^{1 \over 2}
\left({Q \over 10^4}\right)
\left({g \over 10^{-17} {\rm GeV}^{-1}}\right)\cdot\nonumber\\
&\cdot& N_d
\left({{\rm cm} \over r_d}\right)\left({V_m \over {\rm m}^3}\right)
\left({\mu{\rm H} \over L}\right) \left({B_0 \over 10~{\rm T}}\right)\ .
\label{signal}
\end{eqnarray}
In comparison the sensitivity of today's best magnetometers
is $\delta B = B_n \sqrt{\Delta\nu \over {\rm Hz}}$ with 
$B_n$ of order $10^{-16}~{\rm T}$.  A quality factor of 
$10^4$ implies that the detector bandwidth is $10^{-4} \nu$.  
If a factor 2 in frequency is to be covered per year, and the 
duty factor is 30\%, the amount of time spent at each tune of 
the LC circuit is of order $10^3$ seconds.  

The signal to noise ratio will depend on the signal coherence time 
which in turn depends on the velocity dispersion of the axions.  We 
consider two different assumptions for the local axion velocity
distribution.  Assumption A is that the isothermal halo model is 
correct \cite{MST}. In that case the local dark matter density is 
of order $\rho_{\rm dm} \simeq 300$ MeV/${\rm cm}^3$ and the velocity 
dispersion is of order $\delta v \simeq 10^{-3}$.  The energy dispersion 
is of order $\delta E \simeq 10^{-6} m_a$ and hence the coherence time 
$t_c = 1/\delta E \simeq 0.16~{\rm s} ({\rm MHz}/\nu)$ where $\nu$
is the frequency associated with the axion mass: $m_a = 2 \pi \nu$. 
Under assumption A, the magnetometer can detect a magnetic field
$B_d = 10^{-16}~{\rm T}~({\rm Hz})^{-{1 \over 2}}~
(t_c~t)^{-{1 \over 4}} \simeq 2.8 \cdot 10^{-17}~{\rm T} 
\left({\nu \over {\rm MHz}}\right)^{1 \over 4}$ in $t = 10^3$ s of
integration time.  Assumption B is that the caustic ring halo model is 
correct \cite{MWhalo}.  In that case the local dark matter distribution 
is dominated by a single flow with density $\rho_{\rm dm} \simeq$ 
1 GeV/${\rm cm}^3$, velocity $v \simeq$ 309 km/s and velocity 
dispersion $\delta v \lesssim$  53 m/s.  The energy dispersion of 
that flow $\delta E = m_a v \delta v \lesssim 1.8 \cdot 10^{-10} m_a$ 
and hence $t_c \gtrsim 880~{\rm s} ({\rm MHz}/\nu)$.  However, the Earth's 
rotation continually shifts the flow velocity in the laboratory by an 
amount of order 2 cm/s per second.  If this Doppler shift is not 
removed, there is an upper limit on the coherence time of order 
$t_c < 1.4 \cdot 10^3~{\rm s} ({\rm MHz}/\nu)^{1 \over 2}$.  The 
Doppler shift can be partially removed by exploiting information 
about the velocity vector of the locally dominant flow \cite{MWcr}. 
Under assumption B, we expect therefore the signal to be coherent 
over the whole $10^3$ seconds of measurement integration time and 
hence the magnetometer sensivity to be of order $3.2 \cdot 10^{-18}$ T.  
Under assumption B, the signal to noise ratio is approximately a factor 
15 larger than under assumption A, a factor 9 because of the increased 
coherence time and a factor 1.7 because of the increased density.  Recently, 
the caustic ring model has been modified \cite{ban}.  In the modified model, 
the densities of all local flows are increased by a factor of order five.  
The signal to noise ratio is then increased by a factor of order 2.2 
compared to assumption B. 

We now consider other sources of noise, in addition to the noise in the 
magnetometer.  Most importantly, there is thermal (Johnson-Nyquist) noise 
in the LC circuit.  It causes voltage fluctuations 
$~\delta V_T = \sqrt{4 k_B T~R~\Delta \nu}~$ \cite{nyq} and hence current 
fluctuations 
\begin{eqnarray}
&~&\delta I_T = {\delta V_T \over R} = 
\sqrt{4 k_B T Q \Delta\nu \over L \omega}
= 2.96 \cdot 10^{-13} {\rm A} \cdot \nonumber\\
&\cdot&\sqrt{\left({{\rm MHz} \over \nu}\right)
\left({\mu{\rm H} \over L}\right)\left({Q \over 10^4}\right)
\left({T \over {\rm mK}}\right)\left({\Delta \nu \over {\rm mHz}}\right)} 
\label{John}
\end{eqnarray}
where we used the relation $R = {L \omega \over Q}$ between the resistance 
and quality factor of a LC circuit.  We expect that it will be possible to 
cool the LC circuit to below 0.5 ${\rm mK}$ in two stages, using a dilution
refrigerator followed by a nuclear demagnetization refrigerator.  A 
temperature of 0.4 ${\rm mK}$ was achieved at the NHMFL Ultra-High B/T
Facility using this technique \cite{Sull}.  Eq.~(\ref{John}) should be 
compared with the current due to the signal 
\begin{eqnarray}
I &=& {Q \over L} V_m g \dot{a} B_0 = 
1.99 \cdot 10^{-11} {\rm A} \left({Q \over 10^4}\right) 
\left({\mu{\rm H} \over L}\right) \left({V_m \over {\rm m}^3}\right)
\cdot\nonumber\\ &\cdot&
\left({g \over 10^{-17} {\rm GeV}^{-1}}\right)
\sqrt{\rho_a \over {\rm GeV}/{\rm cm}^3}\left({B_0 \over 10~{\rm T}}\right)
\label{sig}
\end{eqnarray}
and with the fluctuations in the measured current due to the noise in the 
magnetometer
\begin{eqnarray}
\delta I_B &\simeq& {2 r_d \over N_d} \delta B = 5.03 \cdot 10^{-14} {\rm A}
\cdot \nonumber\\ &\cdot&
{1 \over N_d} \left({r_d \over {\rm cm}}\right) 
\left({B_n \over 10^{-16}~{\rm T}}\right)
\sqrt{\Delta \nu \over {\rm mHz}}\ .
\label{mag}
\end{eqnarray}
Another possible source of noise is flux jumps in the magnet that produces the 
$\vec{B}_0$ field.  Such flux jumps are caused by small sudden displacements in 
the positions of the wires in the magnet windings.  Since the jumps occur over 
time scales of order $10^{-2}$ to $10^{-3}$ s, the noise they produce at MHz 
frequencies is suppressed.  Such flux jumps are a negligble source of noise 
in ADMX, which however operates at GHz frequencies.  This noise would also 
affect the proposals of ref. \cite{NMR}.  Finally, there are false signals 
associated with man-made electromagnetic radiation.  Such false signals are 
commonly seen in ADMX but can easily be eliminated by various tests.  They 
can be avoided altogether by placing the detector in a Faraday cage but, as 
with ADMX, this may not be necessary.

Assuming that thermal and magnetometer noise are the main backgrounds, 
the signal to noise ratio is
\begin{equation}
s/n = {I \over \sqrt{(\delta_T I)^2 + (\delta_B I)^2}}
\label{son}
\end{equation}
with $I$, $\delta I_T$ and $\delta I_B$ given above, and $L$ given by
Eqs.~(\ref{Lsum}) and (\ref{dind}).  The $s/n$ ratio may be optimized 
with respect to $N_d$ and $r_d$.  It is best to make $r_d$ as small 
as conveniently possible.  The optimal value of $N_d$ is 
\begin{equation}
N_d = \sqrt{{L \over L_e} 
\left(\sqrt{1 + {L_e \over c_d r_d}} - 1 \right)}
\label{opN}
\end{equation}
with 
\begin{eqnarray}
L_e &=& {k_B T Q {\rm Hz} \over r_d^2 B_n^2 \omega} 
= 35 \mu{\rm H} \left({{\rm cm} \over r_d}\right)^2
\left({{\rm MHz} \over \nu}\right) \left({Q \over 10^4}\right)
\cdot \nonumber\\ &\cdot&
\left({T \over {\rm mK}}\right) \left({10^{-16}~{\rm T} \over B_n}\right)^2\ .
\label{Le}
\end{eqnarray}
For the experimental parameters envisaged, the magnetometer noise is 
always much less than the thermal noise.

Fig. II shows the limits that can be placed on $g$ using two specific 
magnets.  In each case, the limits make assumption B for the local axion 
velocity distribution ($t = t_c = 10^3$ s).  Furthermore we assumed 
$Q = 10^4$, $T = 0.5$ mK, and that all axion candidate signals with 
$s/n > 5$ have been ruled out.  The two magnets are: a) the ADMX magnet 
($l_m$ = 1 m, $r_m$ = 0.3 m, $L_m$ = 2.4 $\mu$H, $L_c$ = 0.2 $\mu$H, 
$B_0$ = 8 T), b) the CMS magnet ($l_m$ = 13 m, $r_m$ = 3 m, $L_m$ = 37 $\mu$H, 
$L_c$ = 0.5 $\mu$H, $B_0$ = 4 T).  Because of stray capacitance each LC circuit 
has a maximum frequency.  We calculated the cutoff frequencies assuming that the 
stray capacitance is 15 pF per meter of circuit length.  As discussed above, under 
assumption A for the local axion density and velocity distribution, the expected 
limits are approximately a factor 15 weaker than shown in Fig. II.  

\begin{figure}
\includegraphics[width=0.9\columnwidth]{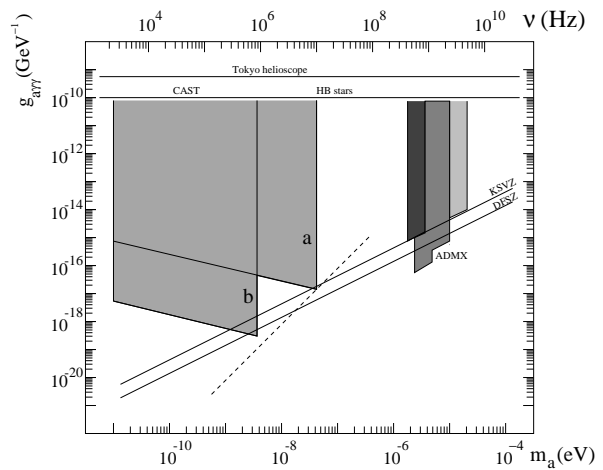}
\caption{Expected senstivity of the proposed detector.
The horizontal lines near the top indicate upper limits
on $g$ from stellar evolution (horizontal branch stars)
and from the Tokyo and CAST solar axion searches \cite{helio}.  
The shaded areas on the right are limits obtained (dark) and 
anticipated (lighter) by the ADMX axion dark matter search.
The light shaded areas on the left show the expected 
sensitivity of the proposed experiment, under the 
assumptions spelled out in the text, using a) the ADMX 
magnet, and b) the CMS magnet.  The dashed line indicates
how the sensitivity scales with the magnet's physical 
size, keeping everything else fixed.}                                                        
\label{lim}                                                     
\end{figure}  

We are grateful to John Clarke for useful comments.
This work was supported in part by the U.S. Department 
of Energy under contract DE-FG02-97ER41029.

\end{document}